\documentclass[a4paper]{iopart}
\usepackage{graphicx}
\newcommand{\ycs}{YbCo$_{2}$Si$_{2}$}
\newcommand{\yrs}{YbRh$_{2}$Si$_{2}$}
\newcommand{\yrcs}{Yb(Rh$_{1-x}$Co$_{x})_{2}$Si$_{2}$}
\newcommand{\et}{~\textit{et al.~}}
\begin{document}
\title{$H~-~T$ phase diagram of \ycs~with $H~//~$[100]}
\author{L Pedrero, M Brando, C Klingner, C Krellner, C Geibel and F Steglich}
\address{Max Planck Institute for Chemical Physics of Solids, N\"othnitzer Str. 40, Dresden 01178, Germany}
\ead{luis.pedrero@cpfs.mpg.de}
\begin{abstract}
We report on the first high-resolution dc-magnetisation ($M$) measurements on a single crystal of \ycs. $M$ was measured down to 0.05~K and in fields up to 12~T, with the magnetic field $H$ parallel to the crystallographic direction [100]. Two antiferromagnetic (AFM) phase transitions have been detected in a field $\mu_{0}H = 0.1$~T at $T_{N} = 1.75$~K and $T_{L} = 0.9$~K, in form of a sharp cusp and a sudden drop in $\chi = M/H$, respectively. These signatures suggest that the phase transitions are $2^{nd}$ order at $T_{N}$ and $1^{st}$ order at $T_{L}$. The upper transition is suppressed by a critical field $\mu_{0}H_{N}~=~1.9$~T. The field-dependent magnetisation shows two hysteretic metamagnetic-like steps at the lowest temperature, followed by a sharp kink, which separates the AFM region from the paramagnetic one. The magnetic $H~-~T$ phase diagram of \ycs~has been deduced from the isothermal and isofield curves. Four AFM regions were identified which are separated by $1^{st}$ and $2^{nd}$ order phase-transition lines.
\end{abstract}
\section{Introduction}
Magnetic ordering in Yb-based intermetallic compounds is quite rare. There are generally two reasons for that: i) In many of these materials, the Yb atom has a valence close to +2, which implies a zero or a very weak $4f$-shell magnetic moment; ii) If the valence is close to +3, a strong hybridisation between the localised $4f$ and the conduction electrons, i.e. the Kondo effect, can progressively screen the Yb magnetic moments and leave a non-magnetic ground state. One of the most studied examples of the latter category is \yrs, where a Kondo temperature $T_{K}\approx 25$~K has been estimated from the magnetic entropy and an antiferromagnetic (AFM) order state has been found just below $T_{N} = 0.07$~K~\cite{trovarelli2000}. In this particular material, the very low transition temperature made it possible to study the unconventional behaviour of the thermodynamic and transport properties caused by a novel type of quantum critical fluctuations~\cite{gegenwart2008}. The real nature of these fluctuations may be observed directly in the magnetic response $S(\textbf{Q},\omega)$, measured by inelastic neutron scattering (INS). These experiments require a certain knowledge of the magneticcally ordered structure below $T_{N}$, e.g., the ordering wave vector $\textbf{Q}$. \\
In \yrs~the structure of the ordered phase is still unknown. This is mainly due to experimental problems: limited size of the single crystals ($V\approx 1$~mm$^{3}$), low $T_{N}$ and also an unexpectedly very small ordered moment~\cite{ishida2003}. To overcome these difficulties, INS experiments should be carried out under pressure, since $T_{N}$ and the corresponding ordered moment increase with increasing pressure in Yb-based compounds~\cite{mederle2001,knebel2006}. However, such experiments have not been successful yet, mainly because of the reasons listed above.\\
Another way to approach the problem is using chemical pressure; isoelectronic substitution of Rh by Co leads to a similar effect as pressure~\cite{westerkamp2008,friedemann2009}. Recently, the crystal growing process has been optimised in order to produce single crystals of \yrcs, which all crystallise in the tetragonal ThCr$_{2}$Si$_{2}$-type structure: Several high-quality single crystals with a Co content $x$ varying between 0.03 and 1 have been synthesised~\cite{klingner2009}. As expected, increasing $x$ stabilises magnetic order, enhancing $T_{N}$ and the value of the ordered moment.\\
The main aim of our study is to investigate the magnetically ordered state in the \yrcs~series, starting with \ycs, and to try to get as much information as possible about the ordered state in \yrs. In this contribution, we present the first low-temperature study of the magnetic $H~-~T$ phase diagram of \ycs, obtained by dc-magnetisation measurements.\\
The first evidence of magnetic order in \ycs~has been observed by~Hodges in $^{170}$Yb M\"ossbauer spectroscopy experiments~\cite{hodges1987}. He found AFM order below 1.7~K with the easy magnetisation along the basal plane and a saturated moment of $1.4~\mu_{B}$/Yb. His results could be explained in terms of a Yb$^{3+}$ valence state experiencing a tetragonal crystalline electrical field (CEF). INS experiments have unambiguously demonstrated that Yb is trivalent at high temperatures and that the Kramers doublet ground state is 4~meV away from the first excited state~\cite{goremychkin2000}.
\begin{figure}[b]
\begin{center}
\includegraphics[width=22pc]{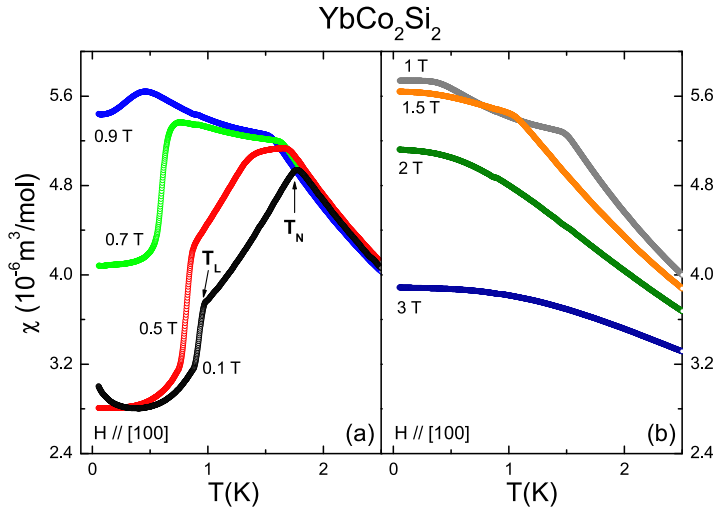}
\end{center}
\caption{\label{fig1} Magnetic susceptibility $\chi = M/H$, plotted as a function of $T$ for small (a) and large (b) fields. $T_{N}$ indicates the upper AFM transition temperature and $T_{L}$ indicates the lower one.}
\end{figure}
This agrees well with our high-temperature susceptibility measurements, which follow a Curie-Weiss behaviour with an effective moment of $4.7~\mu_{B}$/Yb and a Weiss temperature $\theta_{W}$ of -4~K and -160~K for magnetic field $H$ parallel and perpendicular to the basal plane, respectively. No magnetic contribution from Co has been observed~\cite{klingner2009}. A similar fit at temperatures between 2 and 4~K leads to a reduced effective moment of $3.6~\mu_{B}$/Yb. An exact analysis of the CEF level scheme, that might match all previously mentioned results, is needed. Our data will help to solve the CEF Hamiltonian by providing the values of the saturation magnetisation.\\
The sample investigated here has a residual resistance ratio of 2 at 0.35~K and a mass of 35~mg. Its shape is surprisingly square, with sides parallel to the crystallographic direction [100], which made it easy to align the sample. In this paper, our study focuses on measurements with $H~//$ [100]. The dc magnetisation $M$ and susceptibility $\chi = M/H$ has been measured with a high-resolution Faraday magnetometer, in magnetic fields as high as 4~T and temperatures down to 0.05~K~\cite{sakakibara1994}. Owing to the magnetic anisotropy of \ycs~(a factor of about 4~\cite{klingner2009}), the sample platform has been modified so as to reduce the torque contribution to the raw signal, which is proportional to the magnetisation perpendicular to the applied field.
\section{Results}
\begin{figure}[b]
\begin{center}
\includegraphics[width=22pc]{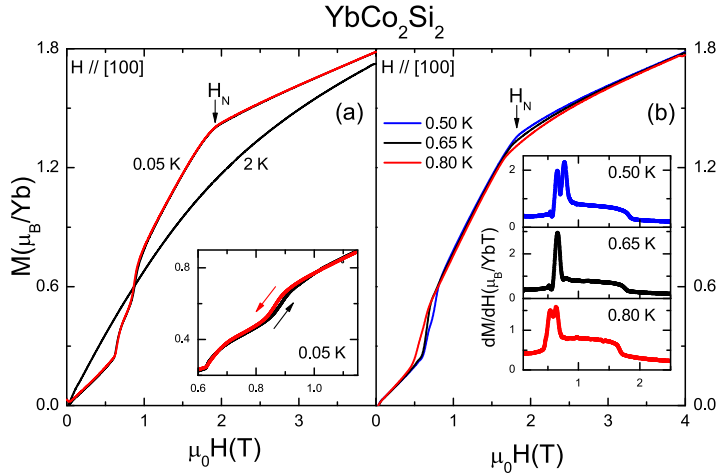}
\end{center}
\caption{\label{fig2} Frame (a): $M$ vs $H$ at the lowest accessible temperature of 0.05~K and at 2~K, just above $T_{N}$. In the inset we zoom on the metamagnetic-like transition to emphasise the hysteresis loop. Frame (b): $M$ vs $H$ at 0.5, 0.65 and 0.8~K. The two metamagnetic-like transitions visible at 0.5~K join each other to become one at 0.65~K and split again at 0.8~K. This feature is accentuated in $dM/dH$ (in form of peaks) plotted in the inset.}
\end{figure}

\begin{figure}[t]
\begin{center}
\includegraphics[width=21pc]{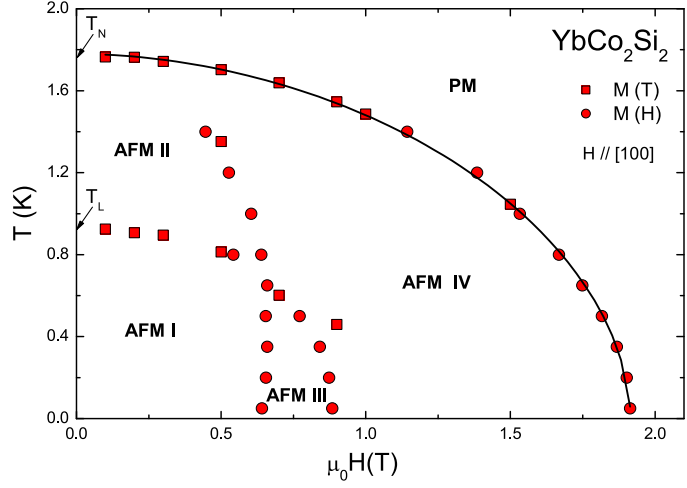}
\caption{\label{fig3} Magnetic phase diagram of \ycs~with $H~//$~[100]. The square and circle points have been extracted from isofield and isothermal measurements of $M$, respectively. PM indicates the paramagnetic region while AFM the antiferromagnatic one. The four different AFM phases are labeled from I to IV. The full line represents the fit performed with the elliptical curve: $[H_{N}(T)/H_{N}(0)]^{n}+[T/T_{N}(0)]^{n} = 1$ where $T_{N}(0)=1.78$~K, $\mu_{0}H_{N}(0)=1.91$~T and $n = 1.9$.}
\end{center}
\end{figure}

Figure~\ref{fig1} shows the temperature dependence of the susceptibility in several fields. At a small field of 0.1~T two features can be seen (indicated by arrows): A sharp kink at $T_{N}=1.75$~K and a distinct drop at $T_{L}=0.9$~K. At $T_{N}$ AFM order sets it and possibly assumes a different AFM structure below $T_{L}$. While the phase transition at $T_{N}$ is $2^{nd}$ order, the sudden drop at $T_{L}$ and the latent heat observed in the heat capacity in zero field indicates the $1^{st}$ order nature of the phase transition; the entropy above both transitions suggests a local character of the Yb $4f$ quasi-hole, pointing to a very low Kondo temperature $T_{K} < 1$~K, compared to that of \yrs~\cite{klingner2009}. Both, $T_{N}$ and $T_{L}$, shift to lower temperatures with increasing $H$. It should be noticed, that at fields higher than 0.1~T, the sharp cusp at $T_{N}$ changes into a plateau where $\chi(T)$ remains almost constant (cf. curve at $\mu_{0}H = 0.5$~T). The phase transition at $T_{N}$ seems to split in field, as can be deduced by the points in the $H~-~T$ phase diagram of figure~\ref{fig3}. The lower transition becomes broader in $T$ as the external field is enhanced above 0.7~T, and, at $H \approx 1.5$~T, disappears. A field of about 2~T is necessary to suppress $T_{N}$ to zero, where $\chi$ becomes nearly constant.
To investigate the phase transition lines in more detail, we measured the field dependence of $M$ at different temperatures. The results are shown in figure~\ref{fig2}. 
The isotherm at 0.05~K shows two metamagnetic-like steps at about 0.65~T and 0.85~T, followed by a kink at $\mu_{0}H_{N} = 1.9$~T, which is associated with the transition at $T_{N}$. Measuring $M$, while sweeping the field up and down, a tiny hysteresis is observed across the first two phase transitions, whereas at $H_{N}$ the transition seems to be continuous. These signatures emphasise the $1^{st}$ and $2^{nd}$ order nature of the phase lines. At a slightly higher temperature $T \approx 0.2$~K, the hysteresis vanishes. By further increasing $T$, the two steps merge into one at 0.65~K and afterwards split again at 0.8~K. This is well illustrated in the inset of figure~\ref{fig2}~(b), where the derivative $dM(H)/dH$ is plotted as a function of $H$. We determined the critical fields of the metamagnetic-like transitions as the fields corresponding to the maxima of the derivative. The points obtained in this way are depicted as circles in figure~\ref{fig3}. The value of the magnetisation just above $H_{N}$ is 1.4~$\mu_{B}/$Yb, which is in agreement with the saturated moment calculated by Hodges~\cite{hodges1987}, and enforces the hypothesis of the local character of the Yb $4f$ quasi-holes. Above $H_{N}$, $M$ increases further, possibly because of the van Vleck contributions to $M$.

\section{$H~-~T$ phase diagram}
The magnetic phase diagram is shown in figure~\ref{fig3}. The squares and circles indicate phase transitions observed in $M$ vs $T$ and $M$ vs $H$, respectively. The outer $2^{nd}$ order phase-transition boundary line, which separates the AFM from the paramagnetic (PM) phase, can be followed from 1.75~K in zero field up to a critical field of 1.9~T, which denotes a potential quantum critical point. The data along this line can well be described by an elliptical curve: $[H_{N}(T)/H_{N}(0)]^{n}+[T/T_{N}(0)]^{n} = 1$ with n = 1.9, $T_{N}=1.78$~K and $\mu_{0}H_{N}(0)=1.91$~T (full line in figure~\ref{fig3}). Inside the magnetic phase, four AFM regions can be identified. Firstof all, the line associated to $T_{N}$ seems to split in field separating the reagion AFM~II and IV by a $2^{nd}$ order phase transition. All other phase boundaries between AFM I and III appear to be $1^{st}$ order line. At first glance, we might interpret our data as follows: Region II is characterised by AFM order with an incommensurable arrangement of the moments, which then assumes a commensurable structure in region I through a $1^{st}$ order phase transition. By applying a magnetic field along [100], the moments arrangement undergoes metamagnetic-like transitions into a canted structure, which only become fully polarised above 1.9~T. \\To extract the magnetic wave vector $\textbf{Q}$ of the ordered phase and its evolution in field, neutron scattering experiments are on the way. For the time being, the magnetic phase diagram for field parallel to all three magnetic non-equivalent crystallographic directions ([100], [110] and [001]) will be presented in a forthcoming paper.\\
\\

\end{document}